\documentclass{article}
\usepackage{amssymb}
\usepackage{bm}
\usepackage[dvips]{graphicx}
\begin{document}

    \title{Current oscillations in a superlattice under non-quantizing electric and magnetic
    fields}

    \author{G. M. Shmelev$^1$, E. M. Epshtein$^2$, M. B.
    Belonenko$^1$\\ \\
    $^1$Volgograd State Pedagogical University, Volgograd, 400131, Russia\\
    $^2$V.A. Kotelnikov Institute of Radio Engineering and Electronics\\
    of the Russian Academy of Sciences, Fryazino, 141190, Russia}
    \date{}
\maketitle

    \abstract{We calculate the current density in a semiconductor superlattice with
    parabolic miniband under crossed non-quantizing electric and magnetic
    fields. The Corbino disk geometry is considered. The current-voltage
    curve contains oscillations with period proportional to the magnetic
    field. The possibility is shown of the negative absolute conductivity.
    The Ampere-Gauss characteristics also contain overshoots under high
    enough electric fields. In all cases, the peaks smear with
    temperature rising.}

\section{Introduction}\label{section1}

In recent experiments, a number of magnetotransport phenomena were found
in solids under classical high magnetic fields in presence of a dc
electric field~\cite{Yang}. In particular, resistance oscillations have been
observed in two-dimensional electron gas (2DEG) under action of both
magnetic and electric fields. Such effects were observed in perfect
GaAs/AlGaAs heterojunctions. A brief review of theoretical and
experimental works on this subject was presented in~\cite{Vavilov}. To interpret the
experimental results unambiguously, further investigations are required of
the nonlinear magnetotransport including the semiconductor structures
other than the heterojunctions mentioned. In this connection, attention
should be paid to~\cite{Gorskii}, where the layered crystal conductivity oscillations
in parallel electric and magnetic fields were studied in scope of the
quasi-classical approach.

In this work, the current behavior is studied under simultaneous effect of
crossed non-quantizing electric and magnetic field on conduction electrons
in a semiconductor superlattice (SL). It should be emphasized here, that a
specific model of a SL with a parabolic miniband (see below) is used. Such
a problem with reference to SLs with cosine miniband has been investigated
by many authors. Some papers should be mentioned in which the electron
motion in electric and magnetic fields was supposed to be non-quantized,
and the electron gas was assumed to be nondegenerate. It was shown in~\cite{Epshtein},
in the linear approximation on the magnetic field, that the Hall field
switching with the sign inversion was possible under high electric field.
In~\cite{Polyanovskii} an exact solution of the electron motion equation and corresponding
current density at zero temperature were found under Corbino regime. No
conductivity oscillations were found there. In~\cite{Bass} the Boltzmann equation
was solved in the $\tau$-approximation and was showed that the current parallel
to SL axis in crossed electric and magnetic fields did not depend on the
magnetic field.

The interest to the miniband models other than cosine one has been revived
in connection with an idea of making a terahertz (THz) Bloch oscillator
based on SL (see, e.g.,~\cite{Suris,Romanov,Shmelev}). As it was shown~\cite{Romanov,Shmelev}, the THz field
generation and amplification conditions can be realized, in particular, in
SL with parabolic miniband. The latter means the electron dispersion law
in form of a truncated parabola (that is the dispersion law is assumed to
be parabolic up to the Brillouin zone edge). Note, that in~\cite{Esaki,Lebwohl} a
dispersion law was considered in form of joined direct and inverted
parabolas (besides the cosine-like law). The parabolic dispersion law in
present work is a specific case of that model.

The magnetic field $\mathbf H$ perpendicular to the driving electric
field $\mathbf E$, which is parallel to SL axis $OX$ transforms an one-dimensional
problem to a two-dimensional one, so that we have, formally, a 2DEG
modulated periodically along SL axis.

\section{The problem statement}\label{section2}

The electron energy in the lowest parabolic miniband of SL is
\begin{equation}\label{1}
  \epsilon(\mathbf p)=\frac{\mathbf p_\bot^2}{2m_\bot}+\varepsilon(p),
\end{equation}
where $\mathbf p_\bot$ and $m_\bot$ are the electron quasi-momentum and
effective mass in the SL layer plane, respectively,
\begin{equation}\label{2}
  \varepsilon(p)=\frac{\Delta d^2}{\pi^2\hbar^2}\frac{p^2}{2},\quad
-\frac{\pi\hbar}{d}\le p\le\frac{\pi\hbar}{d},
\end{equation}
$p$ is the electron quasi-momentum along SL axis, $\Delta/2$ is the
miniband width, $d$ is the SL period. The longitudinal
energy $\varepsilon(p)$ may be expanded into Fourier series:
\begin{equation}\label{3}
  \varepsilon(p)=\frac{1}{2}\sum_{k=1}^\infty\Delta_k\left(1-\cos\left(\frac{kpd}{\hbar}\right)\right),
\end{equation}
where $\Delta_k=4\Delta\displaystyle\frac{(-1)^{k+1}}{k^2\pi^2}$
may be treated as a ``width'' of a partial cosine miniband.

We consider a quasi-classical situation: $\Delta\gg eEd,\,\hbar\omega,\,\hbar/\tau$, where $\omega=eH/(m_\bot c)$ is the
cyclotron frequency, $\tau$ is the mean free time, which is assumed to be
constant.

It is convenient to introduce dimensionless variables by the following substitutions:
\begin{displaymath}
  \frac{d}{\pi\hbar}\mathbf{p}\to\mathbf
p,\quad\frac{t}{\tau}\to t,
  \quad\omega\tau\to\omega,\quad\frac{\mathbf E}{E_0}\to\mathbf E\quad(E_0\equiv\frac{\pi\hbar}{ed\tau}).
\end{displaymath}

To avoid cumbersome expressions, let us make an assumption that is not
principal for the further conclusions, namely, we suupose the longitudinal
effective mass (near the Brillouin zone edge) is equal to the transverse
one, $\displaystyle\frac{\pi^2\hbar^2}{\Delta d^2}=m_\bot$. Such a
condition is fulfilled, e.g., at the following parameter
values: $\Delta=0.02$ eV, $d=5\times10^{-7}$ cm, $m_\bot=10^{-28}$ g. The
electron velocity along the SL axis $OX$ is
\begin{equation}\label{4}
  V_x(t)=\frac{\partial\epsilon(\mathbf p)}{\partial p_x}
  =\frac{2\Delta d}{\pi^2\hbar}\sum_{k=1}^\infty\frac{(-1)^{k+1}}{k}\sin(k\pi p_x(t))
\end{equation}

The time dependence of the quasi-momentum is found from the equation of motion
\begin{equation}\label{5}
  \frac{d\mathbf p}{dt}=\mathbf E+\omega(\mathbf p\times\mathbf h),
\end{equation}
where ~$\mathbf h=\mathbf H/H$ is the unit vector along the magnetic
field.
A situation is considered below, which is realized in the Corbino
geometry,
\begin{displaymath}
  \mathbf E=(E,\,0,\,0),\quad\mathbf H=(0,\,0,\,H).
\end{displaymath}
In this case, the solution of Eq. (\ref{5}) takes the form
\begin{eqnarray}\label{6}
  &&p_x(t)=p_{0x}\cos\omega t+p_{0y}\sin\omega t+\frac{E}{\omega}\sin\omega t,\nonumber\\
&&p_y(t)=p_{0y}\cos\omega t-p_{0x}\sin\omega t+\frac{E}{\omega}(\cos\omega t-1),\nonumber\\
&&p_{0x,\,y}= p_{0x,\,y}(0).
\end{eqnarray}

By Chambers method~\cite{Chambers}, the current density $j$ along the SL axis can be found:
\begin{equation}\label{7}
  j(E,\,\omega,\,T)=e\sum_{\mathbf{p}_0}f_0(\mathbf{p}_0,\,T)\int_0^\infty\exp(-t)V_x(t)\,dt,
\end{equation}
where the equilibrium distribution function of initial momenta of nondegenerate carriers is
\begin{equation}\label{8}
  f_0(\mathbf{p}_0,\,T)=n\left[2\pi T\mathrm{erf}(1/\sqrt{2T})\right]^{-1}
  \exp\left(-\frac{p_{0x}^2+p_{0y}^2}{2T}\right).
\end{equation}
Here $\mathrm{erf}(z)$ is the error function, $n$ is the carrier density,
and the temperature (in energy units) is presented with the substitution
$T/\Delta\to T$. The integration over $p_{0x}$ in Eq.~(\ref{7}) goes from
$-1$ to $1$, while that over $p_{0y}$ from $-\infty$ to $\infty$.

By substituting Eqs.~(\ref{4}) and~(\ref{6}) to Eq.~(\ref{7}), we obtain (in units of $j_0=en\Delta d/(\pi\hbar)$)
\begin{eqnarray}\label{9}
  &&j(E,\,\omega,\,T)=\frac{2}{\pi(1-\exp(-2\pi/\omega))}
\int_0^{2\pi/\omega}\exp(-t)\nonumber\\
&&\times\sum_{k=1}^\infty\nu_k(\omega t,\,T)\frac{(-1)^{k+1}}{k}
\sin\left(\frac{k\pi E}{\omega}\sin\omega t\right)\,dt,
\end{eqnarray}
where
\begin{eqnarray}\label{10}
  &&\nu_k(\omega t,\,T)=\exp(-k^2\pi^2T/2)\nonumber\\
  &&\times\mathrm{Re}(\mathrm{erf}
  (\sqrt{1/2T}(1+ik\pi T\cos\omega t)))/\mathrm{erf}(\sqrt{1/2T}).
\end{eqnarray}
In absence of the magnetic field ($\omega\to0$), Eqs.~(\ref{9})
and~(\ref{10}) lead to the result of~\cite{Shmelev} that was found by solving the
Boltzmann equation in $\tau$-approximation:
\begin{eqnarray}\label{11}
&&j(E,\,0,\,T)=E+\frac{\exp(T/2E^2)}{2\,\mathrm{erf}(1/\sqrt{2T})\sinh(1/E)}\nonumber\\
&&\times\left[\mathrm{erf}(\sqrt{T/2}/E-\sqrt{1/2T})-\mathrm{erf}(\sqrt{T/2}/E+\sqrt{1/2T})\right].
\end{eqnarray}
At $T\to0$, the formula follows from Eq.~(\ref{11}) which was obtained in~\cite{Romanov}:
\begin{equation}\label{12}
   j(E,\,0,\,0)=E-\frac{1}{\sinh(1/E)}.
\end{equation}

\section{The superlattice current-voltage characteristics in magnetic field}\label{section3}
First, let us consider the case of extremely low temperatures ($T\to0$),
when factor~(\ref{10}) is equal to 1, so that Eq.~(\ref{9}) takes the form
\begin{eqnarray}\label{13}
  &&j(E,\,\omega,\,T)=\frac{2}{\pi(1-\exp(-2\pi/\omega))}
\int_0^{2\pi/\omega}\exp(-t)\nonumber\\
&&\times\sum_{k=1}^\infty\frac{(-1)^{k+1}}{k}
\sin\left(\frac{k\pi E}{\omega}\sin\omega t\right)\,dt.
\end{eqnarray}
Note, that the function
$z(p)\equiv\displaystyle\frac{2}{\pi}\sum_{k=1}^\infty\displaystyle\frac{(-1)^{k+1}}{k}\sin k\pi p$
appeared here is defined on the whole number axis and represents the
Fourier expansion of a periodic sawtooth (discontinuous) function, which
is obtained by periodical continuation of the straight-line segment $z=p$
from the interval $-1\le p\le1$. It is such a circumstance that leads,
ultimately, to appearance of the current oscillations. It may be mentioned
that $z(p)$ function often occurs in the theory of magnetic oscillations
in metals~\cite{Shoenberg}.
\begin{figure}
\includegraphics{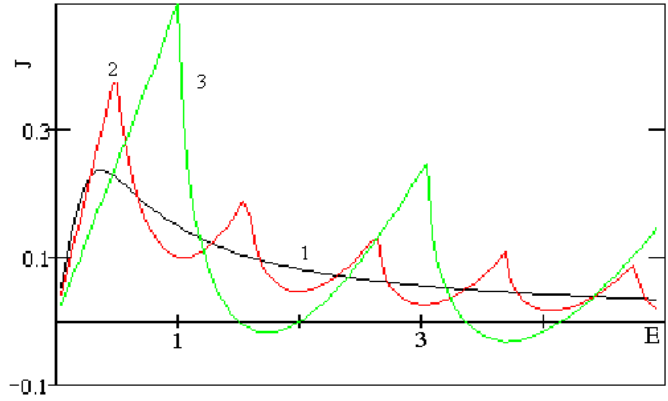}
\caption{Current density as a function of electric field (in
dimensionless units) at $T=0$ and various values of the magnetic field:
1~--- $\omega=0$, 2~--- $\omega=0.5$, 3~--- $\omega=1$}\label{fig1}
\end{figure}

\begin{figure}
\includegraphics{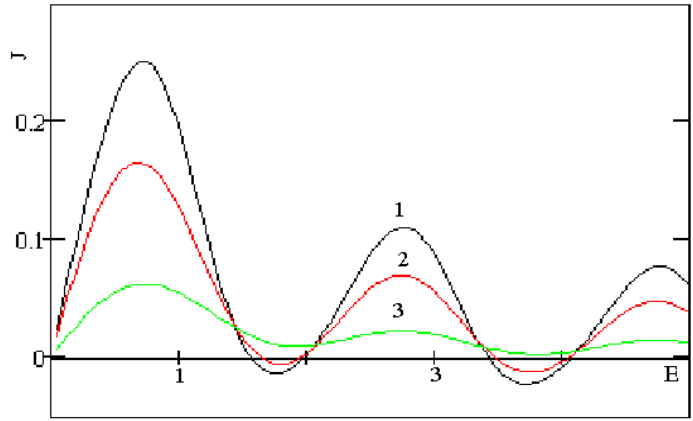}
\caption{Current density as a function of electric field (in
dimensionless units) at $\omega=1$ and various values of the temperature
$T$ (in dimensionless units): 1~--- $T=0.1$, 2~--- $T=0.2$, 3~---
$T=0.5$}\label{fig2}
\end{figure}

\begin{figure}
\includegraphics{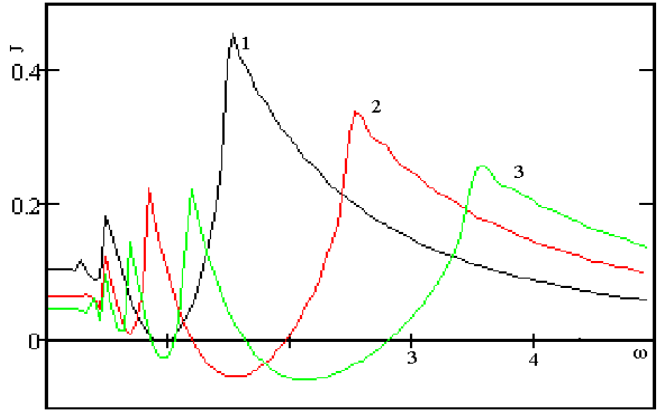}
\caption{Ampere-Gauss characteristics at $T=0$ and various values of
the driving electric field $E$ (in dimensionless units): 1~--- $E=1$,
2~--- $E=2.5$, 3~--- $E=3.5$}\label{fig3}
\end{figure}

Integrating Eq.~(\ref{13}) by parts and using the formula
\begin{equation}\label{14}
  \sum_{k=1}^\infty(-1)^{k+1}\cos kx=\frac{1}{2}-\pi\sum_{n=-\infty}^\infty\delta(x-(2n+1)\pi),
\end{equation}
we get
\begin{eqnarray}\label{15}
  &&j(E,\,\omega,\,0)=\frac{E}{1-\exp(-2\pi/\omega)}\int_0^{2\pi/\omega}\exp(-t)\cos\omega t\nonumber\\
&&\times\left\{1-2\sum_{n=-\infty}^\infty\delta\left(\frac{E}{\omega}\sin\omega t -(2n+1)\right)\right\}\,dt.
\end{eqnarray}
By dividing the integration domain into four equal parts, $\pi/2\omega$
length each, and shifting the variable of integration, we obtain
\begin{eqnarray}\label{16}
  &&j(E,\,\omega,\,0)=\frac{E}{1+\omega^2}-\frac{2}{\cosh(\pi/2\omega)}
\int_0^1\sinh\left(\frac{\pi-2\arcsin x}{2\omega}\right)\nonumber\\
&&\times\left(\sum_{n=0}^\infty\delta(x
-\omega(2n+1)/E)\right)\,dx\qquad\left(\frac{\omega}{E}>0\right).
\end{eqnarray}
Note, that Eq.~(\ref{12}) follows from Eq.~(\ref{16}) at $\omega\to0$. Thus, we get finally from Eq.~(\ref{16})
\begin{eqnarray}\label{17}
  &&j(E,\,\omega,\,0)=j_{0,\,1}\Theta\left(\frac{E}{\omega}\right) \Theta\left(1-\frac{E}{\omega}\right)+
j_{1,\,3}\Theta\left(\frac{E}{\omega}-1\right) \Theta\left(3-\frac{E}{\omega}\right)\nonumber\\
&&+\sum_{s=1}^\infty j_{2s+1,\,2s+3}\Theta\left(\frac{E}{\omega}-(2s+1)\right) \Theta\left(2s+3-\frac{E}{\omega}\right),
\end{eqnarray}
where
\begin{eqnarray}\label{18}
    &&j_{0,\,1}=\frac{E}{1+\omega^2},\nonumber\\
    &&j_{1,\,3}=\frac{E}{1+\omega^2}
-\frac{2\sinh\left(\displaystyle\frac{\arccos(\omega/E)}{\omega}\right)}{\cosh(\pi/2\omega)},\nonumber\\
 &&j_{2s+1,\,2s+3}= j_{2s-1,\,2s+1}\nonumber\\
&&-\frac{2\sinh\left(\displaystyle
\frac{\arccos((2s+1)\omega/E)}{\omega}\right)}{\cosh(\pi/2\omega)},
\end{eqnarray}
$\Theta(x)$ being the Heaviside step function.

\section{Discussion}\label{section4}
Current-voltage curve (CVC) calculated by Eqs.~(\ref{17}) and~(\ref{18})
at $T\to0$ and various values of the magnetic field ($\omega$) is shown in
Fig.~\ref{fig1}. At $\omega\ge0.25$, when the oscillations manifest themselves the most
clearly, the CVC may be called multi-N-type characteristic. Note, that the
magnetic field stimulates the current increasing in the CVC maxima (a
negative magnetoresistance) and leads to appearance of regions with
negative absolute conductivity.

CVC appears overshoots at $E_N=2\omega(N+1/2);\,N=0,\,1,\,\ldots$. The
oscillation period obtained from Eqs.~(\ref{17}) and~(\ref{18}) is
$\Delta(E)=2\omega$, or $\Delta(E)=\displaystyle\frac{2\pi\hbar}{ed}\omega$ (in
dimensional units). At the parameter values mentioned and
$\tau\approx5\times10^{-12}$ s, the electric field unit is $E_0\approx750$
V/cm, while magnetic field $H=10^4$ Oe corresponds to $\omega=1$. Note,
that the quasi-classical conditions are fulfilled at the used parameter
values.
\begin{figure}
\includegraphics{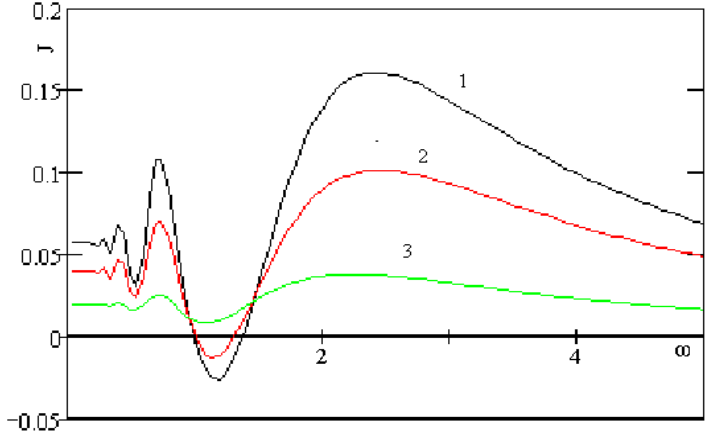}
\caption{Ampere-Gauss characteristics at $E=2$ and various values of
the temperature $T$ (in dimensionless units): 1~--- $T=0.2$, 2~--- $T=0.5$,
3~--- $T=1$}\label{fig4}
\end{figure}
At $T\ne0$ CVC can be calculated numerically by means of Eqs.~(\ref{9})
and~(\ref{10}) (Fig.~\ref{fig2}). In Figs. ~\ref{fig3} and~\ref{fig4} the
Ampere-Gauss characteristics are shown at $T=0$ and $T\ne0$, respectively.
In all the cases, the peaks smear with temperature rising. At the
miniband width value used, $T=1$ value corresponds to 230 K.

Thus, CVC of semiconductor superlattice with parabolic miniband in crossed
classical electric and magnetic fields under Corbino geometry is a
multi-N-type characteristic with oscillations, the period being
proportional to the magnetic field. Appearance of the regions with
negative absolute conductivity is possible. Ampere-Gauss characteristic
contains overshoots. The peaks smear with temperature rising.

\end{document}